\documentclass[aps,prb,twocolumn,showpacs,superscriptaddress,amsmath,amssymb]{revtex4-1}
\usepackage{graphicx}

\begin{document}
\title{Magnetotransport properties of BaRuO$_3$: Observation of two scattering rates}

\author{Y. A. Ying}
\affiliation{Department of Physics and Materials Research Institute, The Pennsylvania State University, University Park, Pennsylvania 16802, USA}

\author{Y. Liu}
\email{liu@phys.psu.edu}
\affiliation{Department of Physics and Materials Research Institute, The Pennsylvania State University, University Park, Pennsylvania 16802, USA}

\author{T. He}
\affiliation{Department of Chemistry, Princeton University, Princeton, New Jersey 08540, USA}

\author{R. J. Cava}
\affiliation{Department of Chemistry, Princeton University, Princeton, New Jersey 08540, USA}

\date{\today}

\begin{abstract}
We report results of low-temperature magnetotransport and Hall measurements on single crystals of four-layered hexagonal (4H) and nine-layered rhombohedral (9R) BaRuO$_3$ that provide insight into the structure-property relationships of BaRuO$_3$ polymorphs. We found that 4H BaRuO$_3$ possesses Fermi-liquid behavior down to the lowest temperature ($T$) of our measurements, 1.8 K. On the other hand, 9R BaRuO$_3$ was found to show a crossover in the temperature dependence of resistivity around 150 K, and the existence of two separate scattering rates at low temperatures. The magnetoresistance in the 9R BaRuO$_3$ was found to be negative while that in the 4H BaRuO$_3$ is positive. We propose that local moments may be present in 9R but not in 4H BaRuO$_3$, which leads to distinctly different behavior in the two forms.
\end{abstract}


\maketitle

Measurements on the Hall angle and resistivity in high-$T_c$ superconductors revealed the rather unusual $\sim T^2$ dependence of the Hall angle and a $\sim T$ dependence of the resistivity. This observation led to the proposal that the transverse scattering rate $\tau_{H}^{-1}$ is different from the transport scattering rate $\tau_{tr}^{-1}$ and a highly unusual charge-spin separation in high-$T_c$ superconductors\cite{CHIEN:1991p246, ANDERSON:1991p264, Coleman:1996p1324}. Similar non-Fermi liquid behavior was also found in other strongly correlated electronic materials such as V$_2$O$_3$\cite{Rosenbaum:1998pR13997} and heavy fermion systems\cite{Paschen:2004p881,Hundley:2004p035113,Nakajima:2004p5}. An interesting question is whether the existence of two scattering rates is limited to strongly correlated electronic systems. It has recently been pointed out that such behavior may also exists in less strongly correlated 4d transition metal oxides such as (Ca,Sr)RuO$_3$ featuring strong multi-orbital correlations\cite{Laad:2008p204}. 

BaRuO$_3$ is a 4d transition metal oxide chemically related to CaRuO$_3$ and SrRuO$_3$ and a striking example of the rich structure-property relationship of oxides. Four different crystalline forms of BaRuO$_3$, known as nine-layered rhombohedral (9R)\cite{Donohue:1965p306,Callaghan:1966p1572,Longo:1968p687}, four-layered hexagonal (4H)\cite{Callaghan:1966p1572,Longo:1968p687}, six-layered hexagonal (6H)\cite{Longo:1968p687}, and cubic perovskite structure (3C), have been synthesized, with the last two forms under extremely high pressures\cite{Jin:2008p197, Zhao:2007p188}. These structures have different amounts of corner and face sharing RuO$_6$ octahedra. The 9R crystallographic form consists of units of three RuO$_6$ octahedra sharing faces in a partial chain, facilitating direct Ru-Ru $d$ orbital interactions within the group. Each of these trimers of octahedra connects to its neighbors along the hexagonal axis by perovskite-like corner sharing with the nearly 180-degree Ru-O-Ru bonds favorable for superexchange coupling. The stacking pattern repeats after nine octahedra. Similarly, the 4H form consists of units of two face-sharing octahedra, connected to the neighbors by corner-sharing, with a repeating stacking pattern along the hexagonal axis after four octahedra. The 6H form consists both dimers of octahedra and single octahedra, arranged alternatively and connected to each other by corner-sharing, forming a more three-dimensional structure. Finally, the 3C form, the most three-dimensional structure in the family, consists of only corner-sharing single octahedra. 

The different structural forms of BaRuO$_3$ lead to different electronic structures\cite{Felser:2000p201} and electrical and magnetic properties. 3C BaRuO$_3$ was found to be ferromagnetic with a transition temperature of 60 K\cite{Jin:2008p197,Zhou:2008p077206}, while 4H and 9R BaRuO$_3$ are both paramagnetic\cite{Rijssenbeek:1999p196,Zhao:2007p188}. 6H BaRuO$_3$, on the other hand, appears to locate at the boundary of ferromagnetic behavior\cite{Zhao:2007p188}. Early transport measurements were carried out on 9R BaRuO$_3$ without identifying the structure-property relationships\cite{Shepard:1997p4978}. Following up measurements on single crystalline 4H and 9R BaRuO$_3$ revealed different temperature dependences of resistivity - the 4H phase is metallic while the 9R phase shows semiconducting behavior at low temperatures\cite{Rijssenbeek:1999p196,Zhao:2007p188,Jin:2008p197}. In addition, results obtained from optical conductivity measurements\cite{Lee:2001p192, Lee:2001p165109} suggest that 4H and 9R BaRuO$_3$ both possess a pseudogap. In this paper, we report results from detailed magnetotransport measurements that provide further insight into the structure-property relationships for the BaRuO$_3$ polymorphs.

Single crystals of 4H and 9R BaRuO$_3$ [Fig. 1 (a) and (b)] used in this study were from the same batches used in the previous study reporting the synthesis and magnetic and electrical characterizations of the materials\cite{Rijssenbeek:1999p196}. The crystals were synthesized by solid-state chemical reactions from powders of BaRuO$_3$, RuO$_2$, and CuO. X-ray diffraction patterns carried out on powders obtained from crushing single crystals showed that the crystals are single-phase. The level of residue copper impurities was checked by energy dispersive x-ray spectroscopy (EDS). No copper was found in the crystals to the sensitivity of an EDS measurement (1-2 $\%$).  The 4H and 9R BaRuO$_3$ crystals used in the present study are small in size ($\sim$ 0.5 mm $\times$ 0.5 mm $\times$ 0.1 mm). Magnetotransport measurements were performed by a standard four-point method. The Hall coefficient was measured by a six-point method. Electrical contacts were prepared in such a way that the current flows in the $ab$-plane. All electrical transport measurements were carried out in a Quantum Design Physical Property Measurement System (PPMS), with a base temperature of 1.8 K. The magnetic fields were applied perpendicular to the $ab$-plane. The magnetoresistance was extracted from the symmetric part of $\rho_{ab}(H)$, while the Hall voltage was extracted from the asymmetric part of $V_{xy}(H)$, from -9 T to 9 T. The Hall voltage was found to be linear as a function of applied field. 

\begin{figure}
\includegraphics[viewport=100 20 510 800,scale=0.48]{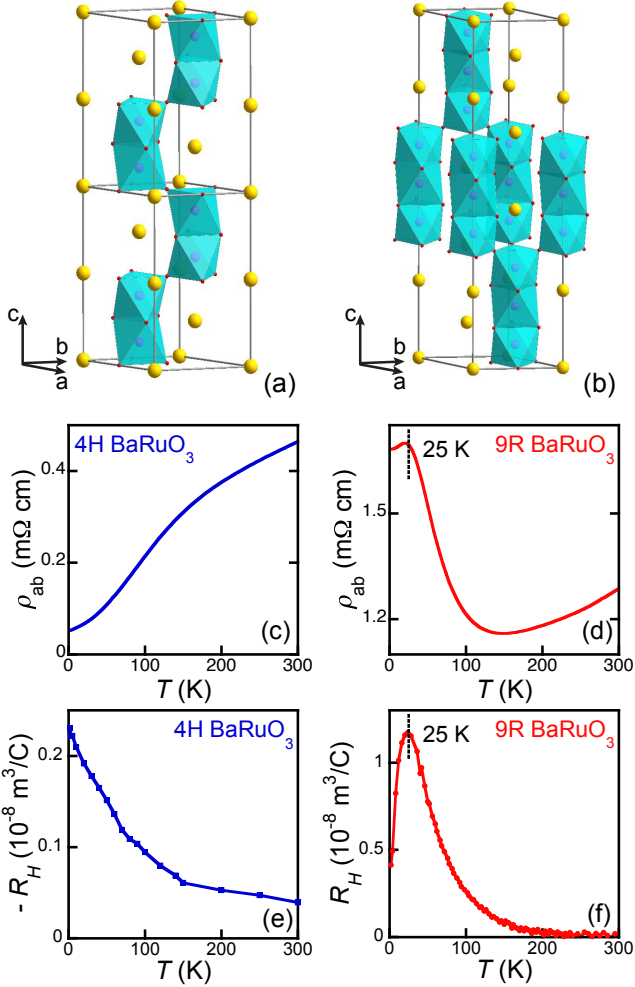}
\caption{(Color online) Crystal structures of (a) 4H BaRuO$_3$ and (b) 9R BaRuO$_3$. Temperature dependence of the in-plane resistivity $\rho_{ab}$ for (c) 4H BaRuO$_3$ and (d) 9R BaRuO$_3$. Temperature dependence of the Hall coefficient $R_{H}$ for (e) 4H BaRuO$_3$ and (f) 9R BaRuO$_3$. A characteristic temperature of $T^{*}$ = 25 K is highlighted in (d) and (f).}
\label{Fig1}
\end{figure}

The electrical transport properties of both 4H and 9R BaRuO$_3$ were found to be similar above 150 K. They display metallic resistivity that follows a linear temperature dependence, as shown in Fig. 1 (c) and (d). The Hall coefficient was found to be weakly temperature dependent for both forms [Fig. 1 (e) and (f)]. The most significant difference between the two forms is the sign of the Hall coefficient, negative for 4H and positive for 9R BaRuO$_3$, indicating the dominance of electron-like carriers in the 4H and hole-like carriers in 9R BaRuO$_3$, consistent with band structure calculations\cite{Felser:2000p201}. 

Below 150 K, the two forms of BaRuO$_3$ showed very different behavior. First, 4H BaRuO$_3$ was found to be metallic down to the base temperature, while 9R BaRuO$_3$ was found to show a crossover from metallic to semiconducting behavior with the resistivity increasing with decreasing temperature [Fig. 1 (c) and (d)]. These findings agree with the previous measurements\cite{Rijssenbeek:1999p196}. Moreover, a saturation in resistivity was observed in 9R BaRuO$_3$ below 25 K, signaling a change in transport scattering rate $\tau_{tr}^{-1}$. The Hall coefficient of 9R BaRuO$_3$ was found to show a maximum also at 25 K, while that of 4H BaRuO$_3$ increases monotonically [Fig. 1 (e) and (f)]. The coincidence between the saturation in resistivity and the maximum in Hall coefficient implies that the electronic state of 9R BaRuO$_3$ may be changing as the temperature is lowered to below 25 K, taken as a characteristic temperature, $T^{*}$. 

\begin{figure}
\includegraphics[viewport=100 200 510 670,scale=0.48]{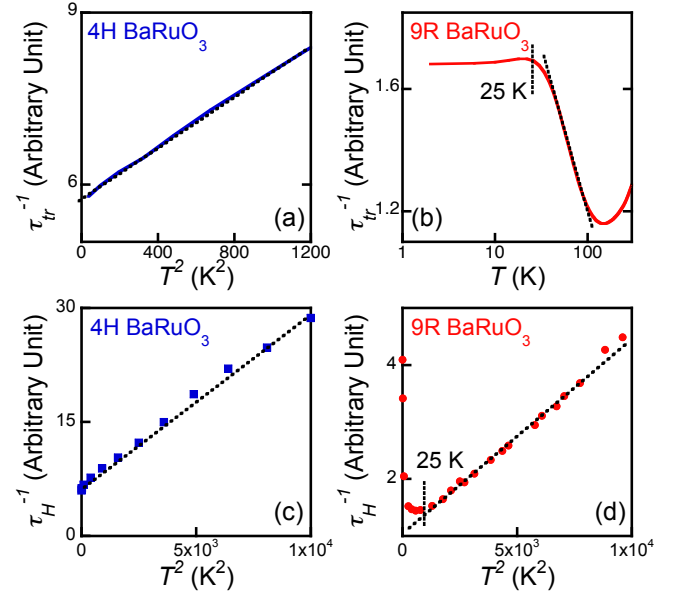}
\caption{(Color online) The transport scattering rate, $\tau_{tr}^{-1} \propto \rho_{ab}$,  for (a) 4H BaRuO$_3$ plotted as a function of $T^2$ and (b) 9R BaRuO$_3$ plotted logarithmically as a function of $T$. The transverse scattering rate, $\tau_{H}^{-1} \propto \cot\theta_{H}$, plotted as a function of $T^2$ for (c) 4H BaRuO$_3$ and (d) 9R BaRuO$_3$. The resistivity saturation coincides with the deviation from $T^2$ dependence of the Hall angle at 25 K in 9R BaRuO$_3$. The dashed lines are guides to the eye.}
\label{Fig2}
\end{figure}

In Fig. 2 (a), we plot the transport scattering rate $\tau_{tr}^{-1}$ calculated from the resistivity $\rho_{ab}$ using the Boltzmann formula as a function of $T^2$ for 4H BaRuO$_3$.  Fermi-liquid behavior, shown by the $\sim T^2$ temperature dependence, was found. In contrast to 4H BaRuO$_3$, the transport scattering rate of 9R BaRuO$_3$ was found to follow a logarithmic temperature dependence $\tau_{tr}^{-1} \propto \ln T$ [Fig. 2 (b)] at low temperatures. Such a temperature dependence may be attributed to the weak localization effects of coherent back scattering by non-magnetic impurities. However, weak localization leads to monotonically increasing resistivity as the temperature is lowered, inconsistent with the saturation observed below $T^{*}$. Alternatively, the logarithmic temperature dependence of resistivity may be explained by Kondo types of scattering by magnetic impurities or local moments.  

The Hall resistivity $\rho_{xy}$ depends on both the transport scattering rate $\tau_{tr}^{-1}$ and the transverse scattering rate $\tau_{H}^{-1}$. The Hall angle is a function of only the transverse scattering rate, $\cot\theta_{H}=\rho_{xx}/\rho_{xy}\propto\tau_{H}^{-1}$. The transverse scattering rate of 4H BaRuO$_3$ was found to display a $\sim T^2$ temperature dependence, the same as the transport scattering rate [Fig. 2 (c)], suggesting the presence of a single scattering rate in 4H BaRuO$_3$ as expected for a standard Fermi-liquid. On the other hand, the transverse scattering rate of 9R BaRuO$_3$ was found to show a $\sim T^2$ dependence down to $T^{*}$, below which the rate increases rapidly [Fig. 2 (d)]. The difference between $\tau_{tr}^{-1}(T)$ and $\tau_{H}^{-1}(T)$ suggests that the scattering in the spin and charge channels are different.

\begin{figure}
\includegraphics[viewport=70 440 530 660,scale=0.98]{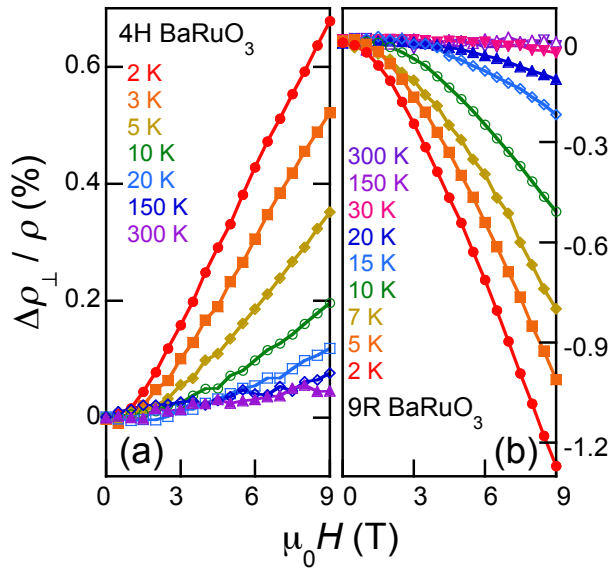}
\caption{(Color online) Transverse magnetoresistance of (a) 4H BaRuO$_3$ and (b) 9R BaRuO$_3$ at various temperatures.}
\label{Fig3}
\end{figure}

The two scattering rates in cuprates were observed below a characteristic temperature well below the onset temperature of the pseudogap\cite{Abe:1999pR15055}. Therefore, even though a pseudogap was indeed observed from optical conductivity measurements in both 4H and 9R BaRuO$_3$ \cite{Lee:2001p192,Lee:2001p165109}, it should be irrelevant to the presence of two scattering rates, similar to the cuprates. We propose that local moments may form in 9R BaRuO$_3$ due to metal-metal bonding, similar to that found in La$_4$Ru$_6$O$_{19}$\cite{Ying:2009p34}. Despite the weak rise seen in the magnetic susceptibility of 9R BaRuO$_3$ below 50 K for field applied along the $c$-axis, which seems not to support the picture of local moments, a small free-moment-like contribution was indeed observed\cite{Rijssenbeek:1999p196}. Such a contribution was interpreted as due to Pauli paramagnetism of the conduction electrons. However, the local moments could cause similar behavior. We speculate that these local moments are absent in 4H BaRuO$_3$ because of the difference in the relative amounts of edge and corner sharing - essentially, 4H BaRuO$_3$ features dimer Ru-Ru bonding while 9R BaRuO$_3$ features trimer Ru-Ru-Ru bonding (see below). In this scenario, the scattering by local moments in 9R BaRuO$_3$ becomes significant below 150 K, giving rise to the $\rho_{ab} \sim \ln T$ behavior above $T^{*}$.

Results obtained from magnetoresistance measurements appear to provide support to the local moment scenario described above. As shown in Fig. 3 (a), the magnetoresistance of 4H BaRuO$_3$ was found to be positive ($\Delta \rho _{\perp} > 0$), which depends quadratically on the applied magnetic field for low fields, consistent with previous results on thin films\cite{Levy:2002p795}. In contrast, the magnetoresistance of 9R BaRuO$_3$ was found to be negative ($\Delta \rho _{\perp} < 0$), rising quickly in magnitude below $T^{*}$. As a magnetic field tends to align the local moments, scattering by local moments is reduced, and negative magnetoresistance is expected. Similar to heavy fermion materials, a coherent state can be enabled by interactions between the local moments below $T^*$, making the Kondo like, individual local moment scattering mechanism invalid. However, a coherent state usually features positive rather than negative magnetoresistance as applying a magnetic field may destroy the delicate coherent screening. Given that a rapid increase in magnetoresistance and a rise in transverse scattering rate occur at similar temperatures, spin scattering must be increasing below $T^{*}$. The saturation of resistivity may be due to the decrease in charge scattering related to the pseudogap opening. 

\begin{table}
\addtolength{\tabcolsep}{3pt}
\caption{Ru-Ru distance of all four forms of BaRuO$_3$, along with several known ruthenates as references.}
\label{Tab1}
\begin{tabular}{c  c  c}
\hline \hline
Material  &  Structure  &  Ru-Ru distance (\AA)  \\[0.5ex]
\hline
La$_4$Ru$_6$O$_{19}$ & cubic & 2.49\\
9R BaRuO$_3$ & rhombohedral & 2.53\\
4H BaRuO$_3$ & hexagonal perovskite & 2.54\\
6H BaRuO$_3$ & hexagonal perovskite & 2.57\\
Bi$_3$Ru$_3$O$_{11}$ & cubic & 2.61\\
Ru metal & hexagonal & 2.65\\
BaRu$_6$O$_{12}$ & hollandite & 2.91\\
3C BaRuO$_3$ & cubic perovskite & 4.00\\ [0.5ex]
\hline \hline
\end{tabular}
\end{table}

Among the four crystalline structures, both 4H and 9R BaRuO$_3$ feature a short distance between Ru atoms in adjacent face sharing RuO$_6$ octahedra ($\sim$ 2.53 \AA), similar to that of La$_4$Ru$_6$O$_{19}$\cite{Ying:2009p34}, much shorter than that of BaRu$_6$O$_{12}$\cite{TORARDI:1985p87}, Bi$_3$Ru$_3$O$_{11}$\cite{Lee:2003p189} and Ru metal (See Table \ref{Tab1}). The short Ru-Ru distance appears to change the Ru $4d$ orbital states, leading to the formation of metal-metal bonding. In the case of La$_4$Ru$_6$O$_{19}$, the strong metal-metal bonding splits the degenerate $t_{2g}$ orbitals, giving rise to molecular-orbital-like states. Such states interact with one another and with the conduction electrons. The competition between the two types of interactions may be responsible for the observed non-Fermi-liquid behavior. Within this picture, the metal-metal bonds within the trimers or dimers of RuO$_6$ octahedra are strong enough to form molecular orbital states, which are formed in such a way that each trimer in 9R BaRuO$_3$ carries a net localized spin, and a local moment. On the other hand, the dimer in 4H BaRuO$_3$ has zero total localized spin. Consequently, no local moments exist in 4H BaRuO$_3$, resulting in weakly positive magnetoresistance. Further measurements and theoretical studies are required to establish this picture.

We would like to thank M. Sigrist, N. Staley, R. Myers and C. Puls for useful discussions and experimental support. The work at Penn State is supported by DOD ARO under Grant No. W911NF-08-1-0220. The work at Princeton is supported by the NSF under Grant No. DMR-1005438.

\end{document}